\documentclass[conference]{IEEEtran}
\IEEEoverridecommandlockouts
\usepackage{multirow}
\usepackage{boldline,multirow}
\usepackage{cite}
\usepackage{algorithm}
\usepackage{graphicx}
\usepackage{textcomp}
\usepackage{xcolor}
\usepackage{booktabs}
\usepackage{tikz}
\usepackage{amsmath}
\usepackage{array} 
\usepackage{subcaption}
\usepackage{bm}
\usepackage{breakurl}
\usepackage{url}
\usepackage{hyperref}
\usepackage{mathtools}
\usepackage[normalem]{ulem}
\usepackage[english]{babel}
\usepackage{amsthm}
\usepackage{amsmath,amssymb,amsfonts}
\usepackage{graphicx,color}
\usepackage{textcomp}
\usepackage{hyperref}
\usepackage{algorithm}
\usepackage{algpseudocode}
\usepackage{placeins}
\usepackage{adjustbox}
\usepackage{amssymb}
\usepackage{bm}
\usepackage{multirow}
\usepackage{threeparttable}
\usepackage{makecell}
\usepackage{bbding}

\usepackage{balance}
\usepackage{textcomp}
\usepackage[T1]{fontenc}

\useunder{\uline}{\ul}{}

\newcolumntype{C}[1]{>{\centering\arraybackslash}p{#1}}
\def\BibTeX{{\rm B\kern-.05em{\sc i\kern-.025em b}\kern-.08em
    T\kern-.1667em\lower.7ex\hbox{E}\kern-.125emX}}
\begin{document}

    \title{HPQEA: A Scalable and High-Performance Quantum Emulator with High-Bandwidth Memory for Diverse Algorithms Support

}

\author{*Hidden for peer-review}
\vspace{-20mm}
\author{
	\IEEEauthorblockN{
    Tran Van Duy\textsuperscript{1},
    Tuan Hai Vu\textsuperscript{1}, 
    Vu Trung Duong Le\textsuperscript{1}, 
    Hoai Luan Pham\textsuperscript{1}, 
    Cao Doanh Bui\textsuperscript{1}, 
    and Yasuhiko Nakashima\textsuperscript{1}}
	\IEEEauthorblockA{
    \textsuperscript{1} Nara Institute of Science and Technology, 8916–5 Takayama-cho, Ikoma, Nara 630-0192, Japan.\\}
Email: tran.van\_duy.tu0@naist.ac.jp
}

\maketitle

\begin{abstract}
    In recent years, there has been a growing interest in the development of quantum emulation. However, existing studies often struggle to achieve broad applicability, high performance, and efficient resource and memory utilization. To address these challenges, we provide HPQEA, a quantum emulator based on the state-vector emulation approach. HPQEA includes three main features: a high-performance computing core, an optimized controlled-NOT gate computation strategy, and effective utilization of high-bandwidth memory. Verification and evaluation on the Alveo U280 board show that HPQEA can emulate quantum circuits with up to 30 qubits while maintaining high fidelity and low mean square error. It outperforms comparable FPGA-based systems by producing faster execution, supporting a wider range of algorithms, and requiring low hardware resources. Furthermore, it exceeds the Nvidia A100 in normalized gate speed for systems with up to 20 qubits. These results demonstrate the scalability and efficiency of HPQEA as a platform for emulating quantum algorithms.

\end{abstract}


\section{Introduction} \label{sec:introduction}
    Recently, quantum computing \cite{nielsen2010quantum} has emerged as a promising solution to complex problems, challenging the limits of Moore’s law \cite{schaller2002moore}. The major technology companies, including IBM, Google, Microsoft, and Fujitsu \cite{SpinQ2025}, are actively developing quantum computers. However, the extreme sensitivity of quantum states confines current systems to controlled laboratory environments. In addition, quantum algorithms play a vital role in exploring applications in fields such as data fitting \cite{wiebe2012quantum}, optimization \cite{wang2024new}, and machine learning \cite{biamonte2017quantum}. However, limited access to practical hardware hinders their development and testing, motivating the use of quantum emulation approaches.
    
    In contrast to actual quantum computers, a quantum bit (qubit) can be emulated on classical computers. Current emulation techniques, such as state-vector-based simulations, are computationally and memory-intensive as the number of qubits (\#Qubits) increases. High-performance computing (HPC) systems, such as DGX A100 and H100 \cite{wang2024queenquickscalablecomprehensive}, can address this issue but introduce extremely high power consumption. In addition, software improvements \cite{10313722} can eliminate unnecessary processing on a small-scale \#Qubits \cite{bergholm2022pennylaneautomaticdifferentiationhybrid}. Other methods, such as tensor networks \cite{10.1007/978-3-031-65633-0_25}, stabilizer formalism \cite{10.1007/978-3-031-65633-0_25}, and density matrices \cite{JOHANSSON20131234}, provide polynomial or linear time computation for specific circuits, but support only limited types of circuits. Therefore, developing an emulation system with high processing performance while minimizing hardware resources and power consumption is still necessary.

    FPGAs are now widely recognized as an efficient platform for quantum emulation \cite{silva2017fpga, mahmud2020efficient, hong2022quantum, waidyasooriya2022scalable, hieu2024, liang2024pcq, vu2024fqsun, tran2025qeaacceleratorquantumcircuit}, enabling deep hardware-level optimization compared to CPUs, GPUs, and HPC. However, existing FPGA-based studies frequently lack thorough hardware optimization \cite{silva2017fpga, mahmud2020efficient, waidyasooriya2022scalable}, support only specific algorithms such as Quantum Fourier Transform (QFT), Quantum Haar Transform (QHT), or Quantum Support Vector Machine (QSVM) \cite{hong2022quantum, liang2024pcq}, use low-memory usage strategies \cite{hieu2024}, or are limited to small-scale systems ($\le$ 17 qubits) \cite{vu2024fqsun, tran2025qeaacceleratorquantumcircuit}.

    To address these limitations, we propose the High-Performance Quantum Emulation Accelerator (HPQEA), which utilizes the state vector-based simulation method to design a hardware architecture that enables deeply optimized computation and storage, high performance with low cost, broad algorithm compatibility, and scalability up to 30 qubits. To gain these goals, HPQEA includes:
        \begin{itemize}
            \item \textbf{High Performance Dual Processing Element Arrays (PEAs):} Accelerate computation and optimize memory usage.
            \item \textbf{Optimized CX Swapper:} Improve data transfer scheduling to reduce CX gate execution latency.
            \item \textbf{Optimized High Bandwidth Memory (HBM) Based Bulk Data Transfer:} Leverage high-bandwidth memory for large-scale data access and storage.
        \end{itemize}

    The AMD Alveo U280 FPGA board was used \cite{AMDAlveoU280} to validate and evaluate HPQEA. The rest of the paper is organized as follows: Section~\ref{sec:background} provides background knowledge; Section~\ref{sec:proposed} explores the hardware architecture of HPQEA; Section~\ref{sec:ver_and_eval} presents implementation and evaluation; and Section~\ref{sec:concl} concludes the study. 

\section{Background knowledge} \label{sec:background}
        \begin{figure*}[t]
            \centering
            \includegraphics[width=1\textwidth]{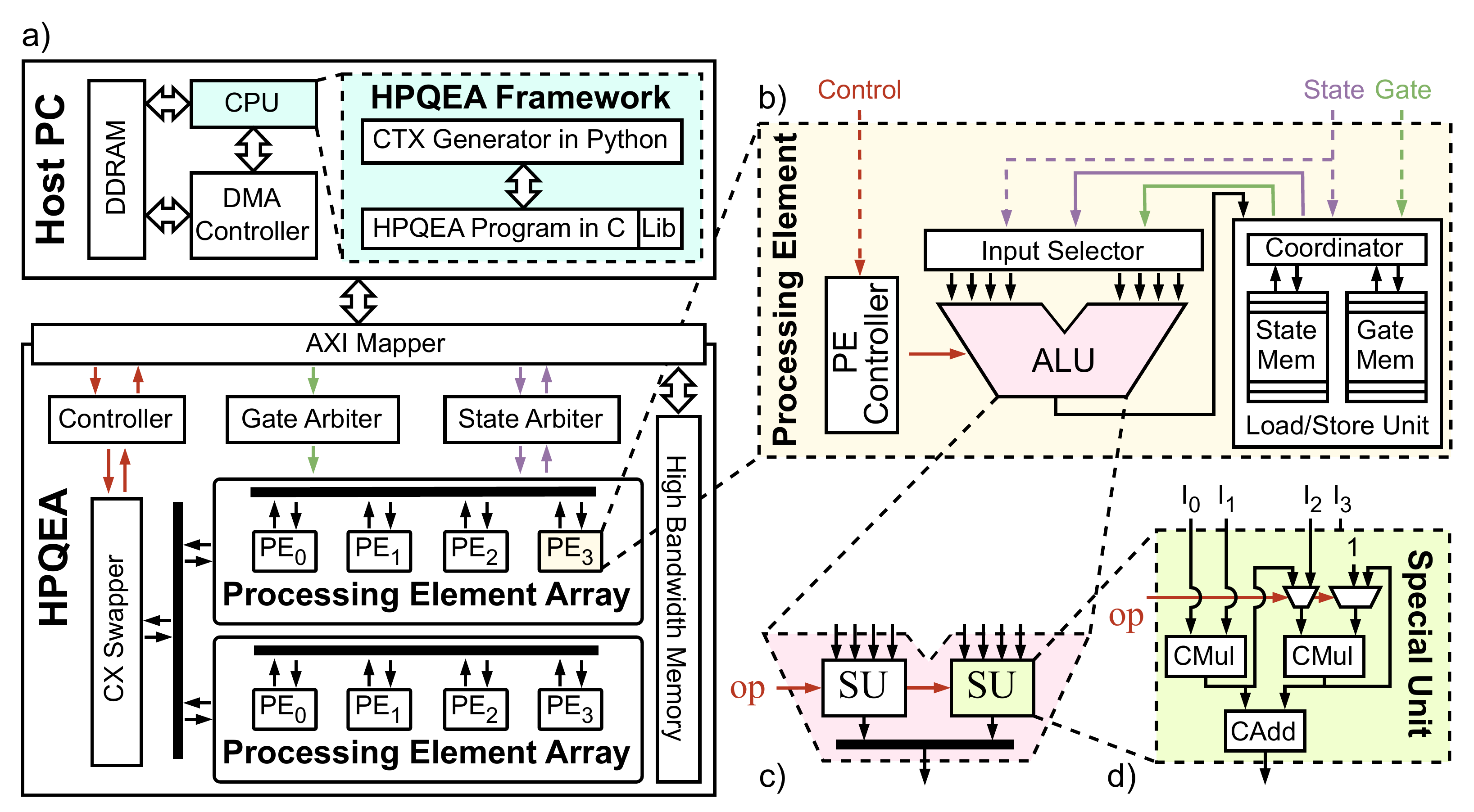}
            \caption{The details of the HPQEA: (a) The overview of the HPQEA system, (b) (Insect) Processing Element, (c) (Insect) Arithmetic Logic Unit, and (d) (Insect) Special Unit.}
            \label{fig:hardware_overview}  
            \vspace{-5mm}
        \end{figure*}
    \subsection{Quantum computing} \label{sec:background:qc_comp}
        Quantum computing leverages fundamental principles of quantum mechanics to process information in ways fundamentally different from classical computing, such as superposition, entanglement, and interference. Generally, the state of an $n$-qubit system at a given time is represented by a $2^n$-dimensional complex vector $|\psi\rangle=\left[\alpha_0\;\alpha_1\;\ldots\;\alpha_{2^n-1}\right]$ where $\{\alpha_j\}$ is called amplitudes.
        
        In addition to the quantum state, quantum circuits update the state vector using quantum gates, which are represented as unitary matrices. In this work, the universal gate set Clifford + $R_i$ with $i\in\{x,y,z\}$ was used following the Gottesman-Knill theorem \cite{PhysRevA.70.052328}. That means any quantum circuit can be presented by the set $\{H, S, R_x(.),R_y(.),R_z(.),CX\}$. A quantum circuit can be broken down into low-level gates or built up into high-level gates. As an example, the control rotation $CR_x(.)$ can be decomposed into $\{S, CX, R_x(.), R_y(.), R_z(.)\}$ following a transpiler \cite{RAKYTA2024112756}. A transpiler can recover the original state by $|\psi_{\text{tran}}\rangle=e^{i\phi}|\psi_{\text{origin}}\rangle$ where $\phi$ is the global phase born from the transpilation process.
    
    \subsection{Quantum simulation algorithm} \label{sec:background:qc_alg}
        
        A quantum circuit is theoretically simulated by applying quantum gates to an initial state $|\psi^{(0)}\rangle$ sequentially. The result state $|\psi^{(m)}\rangle$ after applying a quantum operator (quantum circuit) with $m$ gates is expressed as $|\psi^{(m)}\rangle = \mathcal{U}|\psi^{(0)}\rangle$. The quantum operator $\mathcal{U}$ can be decomposed into a series of matrix multiplications between $p$ smaller sub-operators $u$, as well as a sequence of $g$ quantum gate operations (single-qubit and two-qubit gates) from each sub-operator $u$. These gates can be applied gate by gate to the initial state in sequence, resulting in the same final state as shown in Eq.~\ref{eq:basic}.
        
        \begin{align}
            \mathcal{U} \equiv \prod_{j=1}^p u_j =\prod_{j=1}^p(\bigotimes_{k=1}^{q}g_{k}),\;
            |\psi^{(j+1)}\rangle=u_j|\psi^{(j)}\rangle
            \label{eq:basic}
        \end{align}

\section{Proposed Hardware Architecture}\label{sec:proposed}
    \subsection{Overview} \label{sec:proposed:overview}
        Figure~\ref{fig:hardware_overview} (a–d) shows the overall architecture of the proposed HPQEA system, consisting of two main parts: the host PC (software) and the HPQEA (hardware). On the host PC, the HPQEA framework generates the context data and control signals for the hardware. In detail, the initial quantum state and the quantum circuit context are created using the Python-based Qiskit program, processed by a C program, and transferred to the HPQEA via Direct Memory Access (DMA) over a 256-bit AXI bus for high-throughput communication. The HPQEA hardware includes six main components:
            \begin{itemize}
                \item \textbf{AXI Mapper:} Connects the external AXI bus to the internal memory bus.
                \item \textbf{High Bandwidth Memory:} Provides large storage and high-throughput data access.
                \item \textbf{Controller:} Manages all hardware modules based on host PC commands.
                \item \textbf{Gate Arbiter and State Arbiter:} Distribute quantum state data and gate instructions from the host PC and HBM to internal memory.
                \item \textbf{CX Swapper:} Executes two-qubit CNOT operations.
                \item \textbf{Dual Processing Element Arrays (PEAs):} Update the state vector for single-qubit gate operations.
            \end{itemize}
            
        The following sections provide a detailed discussion of the hardware architecture, including high-performance dual PEAs, optimized CX Swapper, and optimized HBM-based bulk data transfer.

    \subsection{High-Performance Dual PEAs} \label{sec:proposed:high_per_peas}
        Computational performance is critical in quantum simulation, as processing demand grows rapidly with \#Qubits. However, as noted in Section~\ref{sec:introduction}, recent studies often lack deep hardware optimization, broad algorithm support, and efficient resource use. To address these limitations, the proposed HPQEA employs two parallel PEAs, adapted from \cite{tran2025qeaacceleratorquantumcircuit}, effectively doubling the computational ability while preserving wide support for algorithm compatibility and other advantages.

        Each PE (Figure~\ref{fig:hardware_overview}(b)) contains an Arithmetic Logic Unit (ALU) and a Load/Store Unit (LSU). The ALU (Figure~\ref{fig:hardware_overview}(c)) integrates two Special Units (SUs) (Figure~\ref{fig:hardware_overview}(d)) optimized for BRAM bandwidth and computation. These SUs are reconfigurable for sparse or dense single-qubit gate computation \cite{tran2025qeaacceleratorquantumcircuit}, using two complex multipliers, one complex adder, and an operation signal ($op$). The LSU overwrites updated quantum states in the same location as the previous quantum states, avoiding extra memory usage. Moreover, a full $n{\times}n$ quantum gate matrix is replaced with a $2{\times}2$ \cite{tran2025qeaacceleratorquantumcircuit}, reducing resource cost. Unlike QEA \cite{tran2025qeaacceleratorquantumcircuit}, HPQEA utilizes and connects two PEs for data sharing, thereby preventing conflicts and enabling true parallelism.

        Beyond computation, efficient data usage and storage are also important for accuracy and simplified control logic. Following the evaluation in \cite{tran2025qeaacceleratorquantumcircuit}, HPQEA adopts a 32-bit fixed-point format (2 integer bits, 30 fractional bits), providing high fidelity with low mean square error (MSE). To adapt to the dual-PEA structure and support any \#Qubits, the state vector is divided into two parts, one for each PEA, and each part is further divided into four segments for the four PEs in each PEA. The relative positions of state values are preserved without reordering in each PE. This organization allows each PE to process its assigned data independently under two access modes: (1) loading from its own LSU or (2) combining its data and data from another PE. In the second scenario, only the current loaded values of PEs are exchanged over a shared bus, eliminating redundant memory access and avoiding conflicts. This is because the data loaded from the current PE will be the value needed from the other PE, thanks to the proposed data arrangement. This method simplifies data flow control, prevents access conflicts, and improves overall system efficiency.

    \subsection{Optimized CX Swapper} \label{sec:proposed:opt_CX}
        \begin{figure}[t]
            \centering
            \includegraphics[width=0.5\textwidth]{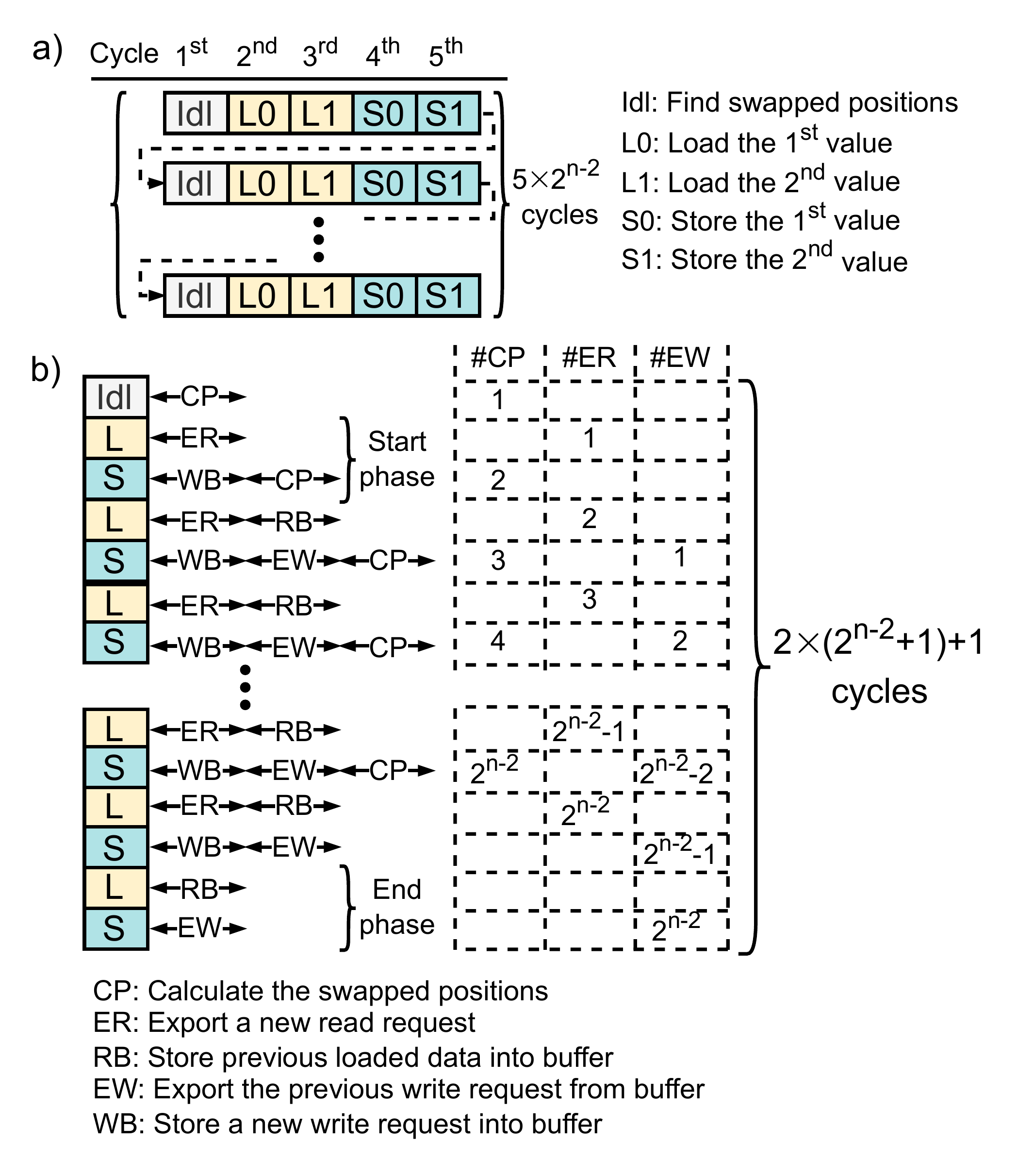}
            \caption{CX working flow: (a) Old version, (b) New version}
            \label{fig:cx_working_flow}  
            \vspace{-5mm}
        \end{figure}
        
        Contrary to other quantum gates, the CX gate just swaps values in the state vector, without performing arithmetic computations \cite{horii2021optimization, tran2025qeaacceleratorquantumcircuit}. However, the design in \cite{tran2025qeaacceleratorquantumcircuit} is inefficient for data scheduling, as illustrated in Fig.~\ref{fig:cx_working_flow}(a). In particular, one cycle is used to compute swap indices, two cycles to load data from the State Memory into the PEs, and two cycles to save the swapped data back. For an $n$-qubit system, this results in $2^{n-2} \times 5$ total cycles, where $2^{n-2}$ represents the number of data pairs swapped. The sequential structure of the process further affects BRAM bandwidth utilization and overall efficiency.

        To address these limitations, the proposed HPQEA implements an improved data schedule, illustrated in Fig.~\ref{fig:cx_working_flow}(b). The operation is organized into three stages: IDLE, LOAD, and STORE. The IDLE stage runs once at the start to compute swap positions for the first paired data. In the LOAD stage, read requests are issued to the State Memory using the precomputed indices to load the swapped data values. During the STORE stage, retrieved data values are written back while the positions for the next swapped values are recalculated. The LOAD and STORE stages repeat until all swaps are completed. Two additional control phases, Start and End, were inserted to ensure correct execution: Start initializes read requests, and End guarantees completion of all write operations. This approach reduces execution time, requiring only $2{\times}(2^{n-2}+1)+1$ cycles, achieving substantial speedup over the baseline sequential design.

    \subsection{Optimized HBM-Based Bulk Data Transfer} \label{sec:proposed:HBM}
        \begin{figure}[h]
            \centering
            \includegraphics[width=0.49\textwidth]{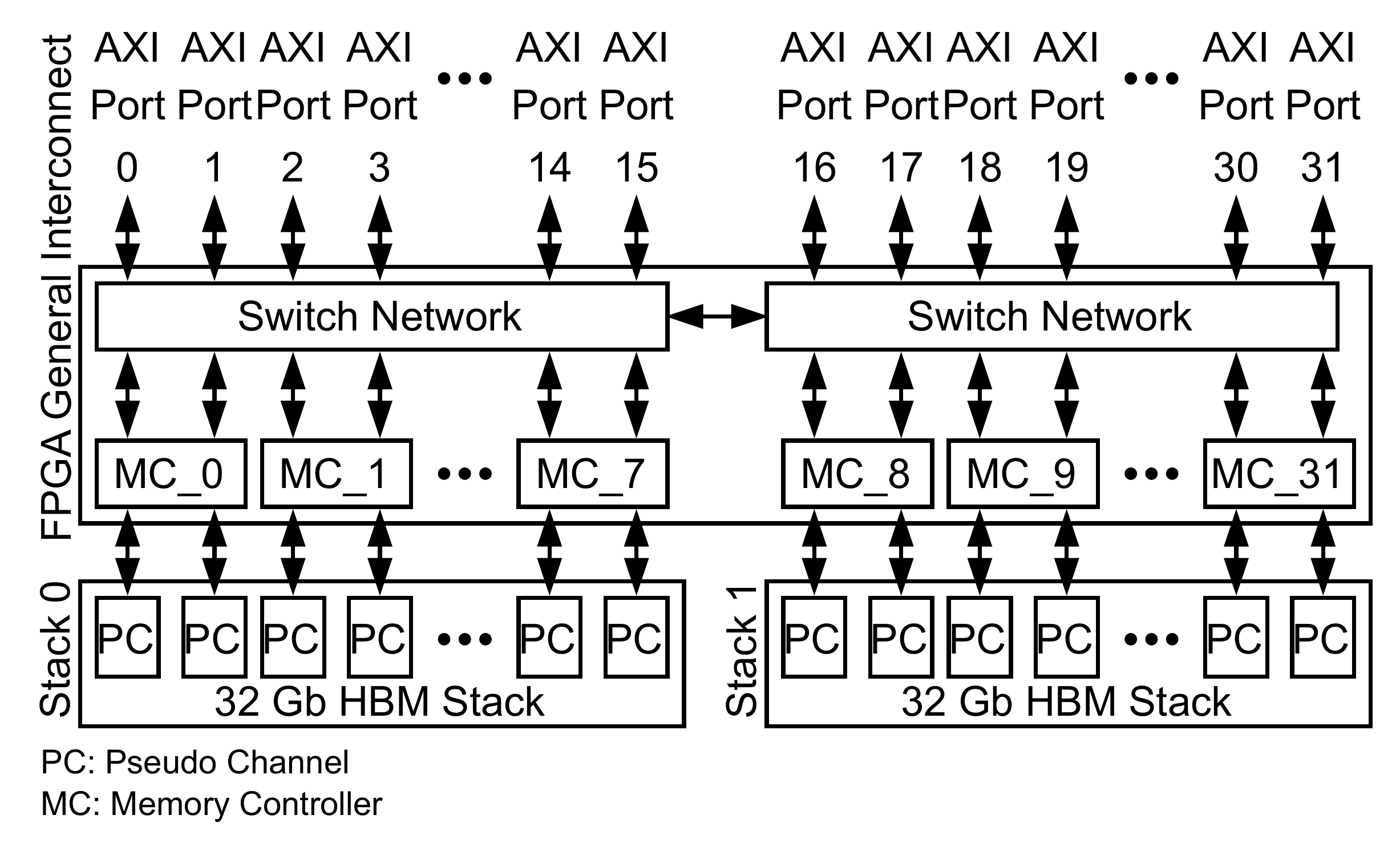}
            \caption{High Bandwidth Memory of the Alveo U280 FPGA board.}
            \label{fig:hbm}  
            \vspace{-5mm}
        \end{figure}
        
        For small-scale systems, such as those with 10 qubits, data transfer is not a critical bottleneck since the state vector can be stored entirely in internal BRAMs. This significantly reduces transfer latency compared to designs that rely on conventional external memory, such as DDR SDRAM. However, as the system scales to higher \#Qubits counts as 25, BRAM capacity becomes insufficient, while DDR SDRAM introduces substantial delays due to limited bandwidth. In this context, HBM offers a more suitable solution. As shown in Fig.~\ref{fig:hbm}, HBM provides more interface ports than DDR SDRAM and is built on more advanced manufacturing technology. Specifically, the Alveo U280 FPGA integrates two HBM stacks, each with a total of 16 AXI ports, offering a combined 32 Gbam of storage. Then, the high number of AXI ports enables HBM to deliver significantly greater bandwidth and lower latency, making it ideal for large-scale quantum simulations.

        With the use of HBM, HPQEA can overcome the scalability limits of the original QEA \cite{tran2025qeaacceleratorquantumcircuit}, which could handle only up to 17 qubits. Although HBM outperforms DDR SDRAM, it still incurs some overhead for initiating and finalizing data transactions via the AXI interface. For systems with fewer than 20 qubits and more than 20 qubits, HPQEA employs a hybrid memory architecture to maximize efficiency. In the first scenario, only internal BRAMs are utilized, thereby eliminating the overhead of external transfers. In the second scenario, the design automatically switches to using HBM. The 19-qubit barrier reflects the resource limitations of the Alveo U280 FPGA used in HPQEA implementations. Then, HPQEA can reduce transfer overhead while effectively scaling any kind of quantum systems.

\section{Verification and Evaluation} \label{sec:ver_and_eval}
    \subsection{Implementation and Verification} \label{sec:ver_and_eval:impl_and_ver}
        To implement and evaluate HPQEA (Fig.~\ref{fig:hardware_overview} (a)), we used the Vivado 2020.2 tool to run the synthesis and implementation on the 16 nm Alveo U280 FPGA board. Table~\ref{tab:utilization} shows the post-implementation results, including the AXI Mapper, Controller, Gate Arbiter, State Arbiter, Dual PEAs, and CX Swapper. In detail, exclusively for HBM, HPQEA requires 43,667 lookup tables (LUTs), 22,634 flip-flops (FFs), 956.5 Block RAMs (BRAMs), and 512 digital signal processors (DSPs), yet it consumes only 0.939 W of power. Additionally, the HBM uses considerably fewer resources, but it consumes 7.4 W due to the management of a large amount of memory (8GB). To ensure the reliability of HPQEA, we also conducted numerous experiments ranging from 3 to 30 qubits, as shown in Section~\ref{sec:ver_and_eval:benchmark:soft_comparison}. In our scenario, 30 is the upper limit due to the memory constraint of the HBM on the Alveo U280 FPGA board. The extensive experiments demonstrated that HPQEA can produce accurate results with very low MSE and high fidelity.

        \begin{table}[h]
            \caption{Post-implementation results of the HPQEA design on the AMD Alveo U280 FPGA}
            \centering
            \setlength{\tabcolsep}{1pt}
            \renewcommand{\arraystretch}{1.3}
            \begin{threeparttable}
            \begin{tabular}{|>{\centering}p{0.11\textwidth}|>{\centering}p{0.05\textwidth}|>{\centering}p{0.055\textwidth}|>{\centering}p{0.045\textwidth}|>{\centering}p{0.05\textwidth}|>{\centering}p{0.058\textwidth}|>{\centering\arraybackslash}p{0.045\textwidth}|}
                \hline
                \multirow{2}{*}{\textbf{Design}} & \textbf{Freq.} &\multicolumn{4}{c|}{\textbf{Resources}} & \textbf{Power} \\
                \cline{3-6}
                & \textbf{(MHz)} & \textbf{LUT} & \textbf{FF} & \textbf{BRAM} & \textbf{DSP} & \textbf{(W)} \\
               \hline
                AXI Mapper & \multirow{8}{*}{250} & 470 & 843 & 0 & 0 & 0.01 \\
                \cline{1-1} \cline{3-7}
                Controller & & 2333 & 695 & 0 & 0 & 0.008 \\
                \cline{1-1} \cline{3-7}
                Gate Arbiter & & 567 & 270 & 28.5 & 0 & 0.011 \\
                \cline{1-1} \cline{3-7}
                State Arbiter & & 1328 & 3420 & 0 & 0 & 0.038 \\
                \cline{1-1} \cline{3-7}
                Dual PEAs & & 27059 & 14811 & 928 & 512 & 0.805 \\
                \cline{1-1} \cline{3-7}
                CX Swapper & & 11910 & 2595 & 0 & 0 & 0.067 \\
                \cline{1-1} \cline{3-7}
                HBM & & 965 & 860 & 0 & 4 & 7.4 \\
                \cline{1-1} \cline{3-7}
                \textbf{HPQEA (Total)} & & \textbf{44632} & \textbf{23494} & \textbf{956.5} & \textbf{516} & \textbf{8.339} \\
                \hline
            \end{tabular}
            \end{threeparttable}
            \label{tab:utilization}
            \vspace{-5mm}
        \end{table}

    \subsection{Benchmarking Methods} \label{sec:ver_and_eval:benchmark}
\begin{figure}[h]
            \centering
            \includegraphics[width=0.49\textwidth]{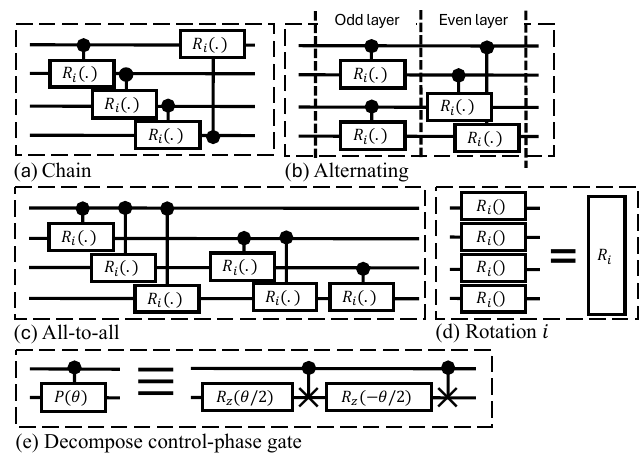}
            \caption{Example of 4-qubit topologies (a-d) and (e) Decomposing of $CP(\theta)$ as a combination of $CX$ and $R_z$ gates \cite{tran2025qeaacceleratorquantumcircuit}.}
            \label{fig:topology}  
        \end{figure}

        To demonstrate the scalability, efficiency, and flexibility of HPQEA across diverse scenarios, we conducted evaluations using 19 quantum circuit templates \cite{expressibility}, each assigned a unique circuit ID from 1 to 19. Each circuit includes various common quantum circuit topologies, including chain, alternating, all-to-all \cite{PRXQuantum.2.040309}, and rotation-based structures, as illustrated in Fig.~\ref{fig:topology}. In detail, the chain topology arranges qubits linearly, allowing interactions only between neighboring qubits. The alternating topology forms a bipartite structure, enabling interconnections between two distinct subsets of qubits. The all-to-all topology represents a fully connected graph, where every qubit can interact directly with all others. Lastly, the rotation $i$ serves as a parameterized single-qubit operation without entanglement.

        \begin{figure*}[t]
            \centering
            \includegraphics[width=0.995\textwidth]{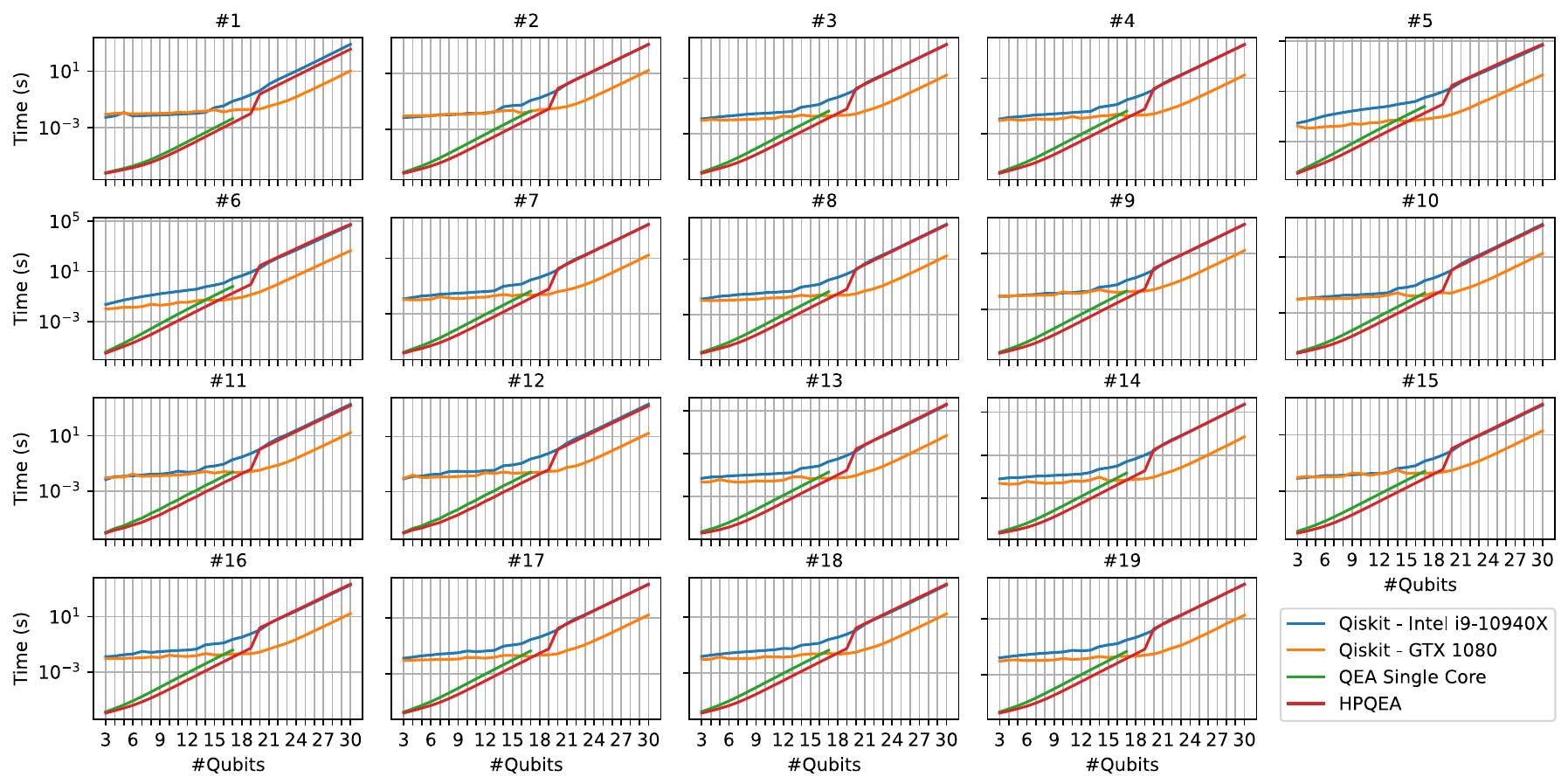}
            \caption{Comparison in execution time between HPQEA and a variety of random quantum circuits indexed from \#1 to \#19}
            \label{fig:execution_time_quanv_comparison}  
            \vspace{-5mm}
        \end{figure*}  
        
        \begin{figure}[t]
            \centering
            \includegraphics[width=0.495\textwidth]{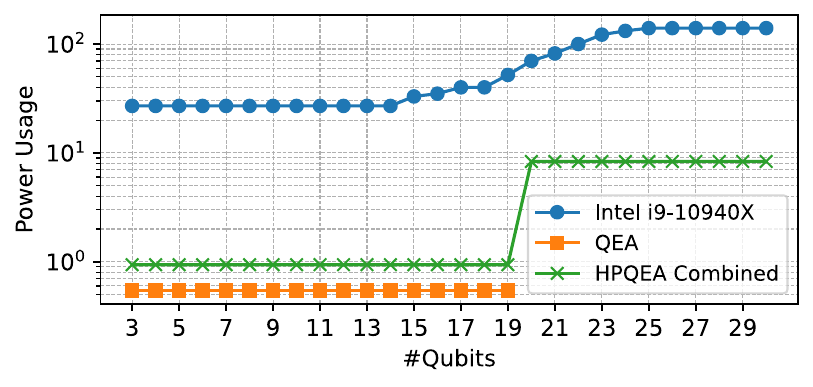}
            \caption{The average power usage comparison between HPQEA and other platforms of random quantum circuits indexed from \#1 to \#19}
            \label{fig:power_usage}  
            \vspace{-5mm}
        \end{figure}  

        To compare with related works, we also evaluated the HPQEA using the Quantum Fourier Transform (QFT) \cite{coppersmith2002approximatefouriertransformuseful}. QFT is an essential part used in various quantum algorithms, such as phase estimation, order-finding, and Shor's factoring algorithm. In our implementation, QFT circuits were generated utilizing Hadamard ($H$) gates, decomposed controlled phase ($CP(\theta)$) in Fig.~\ref{fig:topology} (e), and SWAP operations using three consecutive CNOT ($CX$) gates.

        
        The evaluated metrics include memory usage, fidelity, mean square error (MSE), and execution time, clearly defined in \cite{vu2024fqsun}. The fidelity and MSE between two state vectors range from 0 to 1, with a fidelity of 1 or an MSE of 0 meaning that two vectors completely overlap and vice versa. The execution time on the CPU (Qiskit) is measured from constructing the circuit to receiving the final state, while the execution time of HPQEA is measured by running HPQEA with a maximum frequency of 250 MHz to calculate the final state from the initial state. 
 
        \begin{figure}[t]
            \centering
            \includegraphics[width=0.495\textwidth]{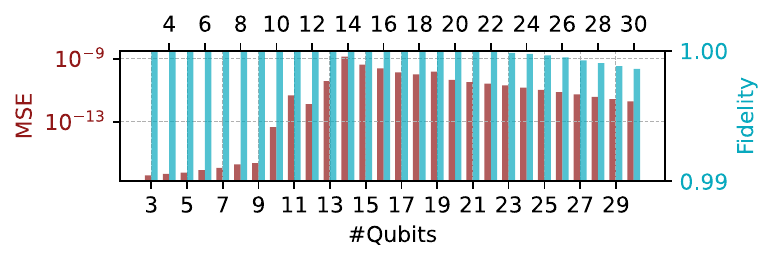}
            \caption{Average MSE and fidelity measurement between HPQEA and Qiskit for a set of templates indexed from 1 to 19.}
            \label{fig:mse_fidelity}  
            \vspace{-5mm}
        \end{figure}

        Furthermore, the evaluation considers four key metrics: memory usage, fidelity \cite{vu2024fqsun}, mean-square error (MSE), and execution time. Fidelity and MSE are used to assess the similarity between two quantum state vectors. Both metrics are normalized between 0 and 1, where a fidelity of 1 or an MSE of 0 indicates perfect overlap between vectors, while lower fidelity or higher MSE indicates increasing divergence. Execution time is measured differently for software and hardware platforms. For the software-based evaluation (Qiskit), the total execution time includes circuit construction, simulation, and retrieval of the final state vector. Besides, the hardware execution time is measured by starting the HPQEA to calculate the final state from the initial state with a fixed maximum operating frequency of 250 MHz.

        \begin{table*}[h]
    		\centering
    		\renewcommand{\arraystretch}{1.2}
    		\resizebox{0.99\linewidth}{!}{\begin{threeparttable}
            \caption{Comparative analysis in post-implementation of HPQEA and existing FPGA-based emulators on QFT's performance.}
            \label{tab:qft}
            \begin{tabular}{|c|c|c|c|c|c|c|c|c|}
                \hline
                \textbf{Works} &
                \textbf{Device} & \textbf{Freq (MHz)}
                & \textbf{Reconfig$\star$} &
                \textbf{Precision} &
                \textbf{\#Qubits} &
                \begin{tabular}[c]{@{}c@{}}\textbf{Execution}  \textbf{time (s)}\end{tabular} &  
                \textbf{\#Gates $^{\dagger}$} &
                \textbf{NGS $^{\dagger\dagger}$} \\ 
                \hline
                \cite{silva2017fpga} & \begin{tabular}[c]{@{}c@{}}AMD Xilinx\\ Zynq-7000 \end{tabular} & 100 & \Checkmark & 32-bit FX & 6 & $1.15\times 10^{-4}$ & 10 & $1.8\times 10^{-7}$ \\ 
                \hline
                \cite{mahmud2020efficient} & \begin{tabular}[c]{@{}c@{}}Arria\\ 10AX115N4F45E3SG\end{tabular} & 233 & \Checkmark & 32-bit FP & 16 & $1.84\times10^{1}$ & 528 & $5.33\times 10^{-7}$ \\ 
                \hline
                \cite{hong2022quantum} & \begin{tabular}[c]{@{}c@{}}Xilinx XCKU115 \end{tabular} & 160 & \XSolidBrush & 16-bit FX & 16 & $2.70\times 10^{-1}$ & 136 & $3.03\times 10^{-8}$ \\ 
                \hline
                \cite{waidyasooriya2022scalable}& \begin{tabular}[c]{@{}c@{}} 2 $\times$ Intel Stratix 10\\ MX2100 \end{tabular} & 299 & \XSolidBrush & 32-bit FP & 30 & $4.47\times 10^0$ & 465 & $8.95\times 10^{-12}$ \\ 
                \hline
                \cite{liang2024pcq} & Xilinx XCVU9P & 233 & \XSolidBrush & 18-bit FX & 16 & $1.20\times 10^{-3}$ & - & - \\ 
                \hline
                \cite{vu2024fqsun} & Xilinx ZCU102 & 125 & \Checkmark & 32-bit FX & 17 & $2.81 \times 10^{0}$ & 721 & $2.97 \times 10^{-8}$ \\ 
                \hline
                \cite{tran2025qeaacceleratorquantumcircuit} & AMD Alveo U280 & 250 & \Checkmark & 32-bit FX & 17 & $3.29\times 10^{-1}$ & 721 & $3.48\times 10^{-9}$ \\ 
                \hline
                \textbf{This work} & \textbf{AMD Alveo U280} & \textbf{250} & \Checkmark & \textbf{32-bit FX} & \textbf{17} & $\bm{9.66\times 10^{-2}}$ & \textbf{721} & $\bm{1.02\times 10^{-9}}$ \\ 
                \hline
            \end{tabular}
            \begin{tablenotes}
                \item[$^{\star}$] The quantum emulator is fixed with a application and \#Qubits. 
                \item[$^{\dagger}$] The $\#\text{Gate}$ in this work is higher than other work due to the no-use of control-rotation gates.
                \item[$^{\dagger\dagger}$] The \textbf{N}ormalized \textbf{G}ate \textbf{S}peed (NGS) (s / (gate $\times$ amplitude)) = Execution time / (\#Gates $\times$ $2^{\#\text{Qubits}}$), smaller is better.
            \end{tablenotes}
            \label{tab:hardware_compare}
    		\end{threeparttable}}
            \vspace{-5mm}
    	\end{table*}
    \subsection{Comparison with CPUs and GPUs} \label{sec:ver_and_eval:benchmark:soft_comparison}
        To validate the correctness and performance, we compared HPQEA with Intel(R) Core(TM) i9-10940X, NVIDIA GTX 1080 (using Qiskit-GPU~\cite{javadiabhari2024quantumcomputingqiskit}), and QEA~\cite{tran2025qeaacceleratorquantumcircuit} across quantum circuits, mentioned in Section~\ref{sec:ver_and_eval:benchmark}, indexed from \#1 to \#19 in Fig.~\ref{fig:execution_time_quanv_comparison}. The results indicate that HPQEA surpasses Intel i9-10940X CPU for systems with less than 19 qubits, with an average acceleration of $\textbf{4}\times$ over QEA. HPQEA also significantly outperforms the GTX 1080 between 3 and 18 qubits systems, but has limited processing capabilities and HBM overhead in systems over 20 qubits.

        In addition, power analysis also highlights HPQEA’s efficiency in Fig.~\ref{fig:power_usage}. While it consumes slightly more power than QEA $\textbf{1.73}\times$ ($0.939$ .vs $0.543$ W) but achieves significantly faster execution times. Furthermore, as compared to the Intel i9-10940X CPU, HPQEA helps save from $\textbf{28.75}\times$ ($0.939$ vs $27$ W) and $\textbf{149.1}\times$ ($0.939$ vs $140$ W). Fidelity and MSE values, presented in Fig.~\ref{fig:mse_fidelity}, are almost identical to Qiskit's results, indicating correctness and numerical reliability.

\begin{figure}[h]
            \centering
            \includegraphics[width=0.495\textwidth]{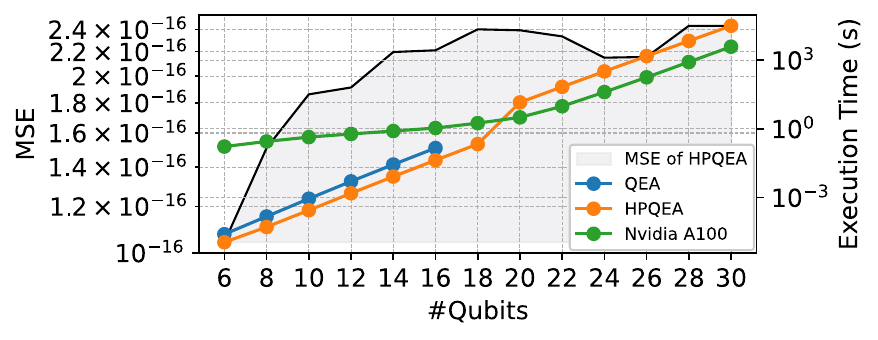}
            \caption{Comparison of QFT circuits in terms of mean square error (MSE) between Qiskit and HPQEA state vectors (Black line, left-y axis), and the execution time between HPQEA and other platforms (right y-axis)}
            \label{fig:qft}  

        \end{figure}

        For a more meaningful benchmark, Fig.~\ref{fig:qft} compares execution times QFT circuits across HPQEA, QEA, and NVIDIA A100. Compared to QEA, it delivers speedups between $\textbf{2.24}\times$ and $\textbf{3.39}\times$, while outperforming A100 by up to $\mathbf{10^4 \times}$ at quantum circuits below 20 qubits. Beyond 20 qubits, however, HPQEA’s performance decreases, running between $\textbf{4.5}\times$ and $\textbf{8.44}\times$ slower than A100. The slowdown is caused by limited computational resources and HBM overhead. This bottleneck could be addressed by increasing computing resources and optimizing HBM usage. In addition, the MSE results remain consistent with previous experiments, verifying the reliability of HPQEA.

    \subsection{Comparison with FPGA-based works} \label{sec:ver_and_eval:FPGA_comparation}    
        To provide a more comprehensive assessment of the efficiency and optimization of HPQEA, this section presents a comparative evaluation against FPGA-based quantum accelerators using the QFT algorithm. 
        
        Table~\ref{tab:hardware_compare} first summarizes the hardware characteristics of HPQEA alongside prior works \cite{liang2024pcq, mahmud2020efficient, hong2022quantum, waidyasooriya2022scalable, silva2017fpga, vu2024fqsun, tran2025qeaacceleratorquantumcircuit}, focusing on key metrics such as maximum operating frequency, reconfigurability, supported \#Qubits, and Normalized Gate Speed (NGS). The NGS metric enables fair comparisons across platforms with differing hardware resources, precision levels, and quantum circuit sizes. For consistency across evaluations, we used a 17-qubit QFT circuit as a reference benchmark, which is also supported by the comparative designs. The results demonstrated that HPQEA achieves superior performance and flexibility compared to the other works. Specifically, HPQEA outperforms \cite{liang2024pcq, mahmud2020efficient, hong2022quantum, silva2017fpga, tran2025qeaacceleratorquantumcircuit, vu2024fqsun} in terms of maximum frequency, with improvements ranging from \textbf{1.07}$\times$ (250 vs. 233 MHz) to \textbf{2.5}$\times$ (250 vs. 100 MHz). Furthermore, HPQEA supports a wider range of quantum algorithms and larger \#Qubits systems, enhancing its adaptability for diverse applications. In terms of NGS, HPQEA exhibits substantial improvements over earlier designs, achieving speedups from \textbf{3.41}$\times$ ($1.02 \times 10^{-9}$ vs. $3.48 \times 10^{-9}$ seconds) up to \textbf{522.55}$\times$ ($1.02 \times 10^{-9}$ vs. $5.33 \times 10^{-7}$ seconds). In comparison with \cite{waidyasooriya2022scalable}, HPQEA could not surpass it in every individual metric, such as operational frequency, supported qubits, or NGS. However, it still offers a significantly wider range of supported algorithms since it is not limited to QFT. In summary, these results highlight HPQEA’s superior performance, configurability, and scalability when compared to existing FPGA-based quantum accelerators.
        
\section{Conclusion} \label{sec:concl}
    This study introduces a high-performance quantum emulation accelerator (HPQEA) designed to address challenges in terms of broad applicability, scalability, performance, and resource efficiency for emulating quantum circuits. HPQEA incorporates three fundamental concepts: a high-performance computing core, an optimized controlled-NOT gate computation strategy, and effective utilization of high-bandwidth memory. Experimental results have shown that it has a wide range of applications, is highly reliable, and uses resources efficiently, outperforming CPU, GPU, and FPGA-based techniques by supporting up to 30 qubits. Nonetheless, execution times were impacted by restricted computational resources and inefficient HBM usage. The future study will focus on optimizing HBM usage while discovering efficient resource utilization strategies to address these constraints.

\section*{Acknowledgment}
This work was supported by JST-ALCA-Next (JPMJAN23F4), the NAIST Senju Monju Project (Daiichi-Sankyo ”Habataku” Support Program for the Next Generation of Researchers).

\bibliographystyle{IEEEtran}
\bibliography{references.bib}

\end{document}